\documentclass[preprint,superscriptaddress,aip,onecolumn]{revtex4-1}
\usepackage{graphicx}
\usepackage{dcolumn}
\usepackage{bm}
\usepackage{amsmath}
\usepackage[english]{babel}
\usepackage{hyperref}
\usepackage{color}

\begin{document}

\title{Movable High-$Q$ Nanoresonators Realized by Sub-wavelength III/V Semiconductor Nanowires on a Si Photonic Crystal Platform}
\author{Muhammad Danang Birowosuto}
\affiliation{NTT Basic Research Laboratories, NTT Corporation, 3-1 Morinosato Wakamiya, Atsugi, Kanagawa 243-0198, Japan}
\affiliation{NTT Nanophotonics Center, NTT Corporation, 3-1 Morinosato Wakamiya, Atsugi, Kanagawa 243-0198, Japan}
\author{Atsushi Yokoo}
\affiliation{NTT Basic Research Laboratories, NTT Corporation, 3-1 Morinosato Wakamiya, Atsugi, Kanagawa 243-0198, Japan}
\affiliation{NTT Nanophotonics Center, NTT Corporation, 3-1 Morinosato Wakamiya, Atsugi, Kanagawa 243-0198, Japan}
\author{Guoqiang Zhang}
\affiliation{NTT Basic Research Laboratories, NTT Corporation, 3-1 Morinosato Wakamiya, Atsugi, Kanagawa 243-0198, Japan}
\author{Kouta Tateno}
\affiliation{NTT Basic Research Laboratories, NTT Corporation, 3-1 Morinosato Wakamiya, Atsugi, Kanagawa 243-0198, Japan}
\author{Eiichi Kuramochi}
\affiliation{NTT Basic Research Laboratories, NTT Corporation, 3-1 Morinosato Wakamiya, Atsugi, Kanagawa 243-0198, Japan}
\affiliation{NTT Nanophotonics Center, NTT Corporation, 3-1 Morinosato Wakamiya, Atsugi, Kanagawa 243-0198, Japan}
\author{Hideaki Taniyama}
\affiliation{NTT Basic Research Laboratories, NTT Corporation, 3-1 Morinosato Wakamiya, Atsugi, Kanagawa 243-0198, Japan}
\affiliation{NTT Nanophotonics Center, NTT Corporation, 3-1 Morinosato Wakamiya, Atsugi, Kanagawa 243-0198, Japan}
\author{Masato Takiguchi}
\affiliation{NTT Basic Research Laboratories, NTT Corporation, 3-1 Morinosato Wakamiya, Atsugi, Kanagawa 243-0198, Japan}
\affiliation{NTT Nanophotonics Center, NTT Corporation, 3-1 Morinosato Wakamiya, Atsugi, Kanagawa 243-0198, Japan}
\author{Masaya Notomi}
\affiliation{NTT Basic Research Laboratories, NTT Corporation, 3-1 Morinosato Wakamiya, Atsugi, Kanagawa 243-0198, Japan}
\affiliation{NTT Nanophotonics Center, NTT Corporation, 3-1 Morinosato Wakamiya, Atsugi, Kanagawa 243-0198, Japan}
\date{\today}

\baselineskip24pt
\begin{abstract}
Sub-wavelength semiconductor nanowires have been attracting strong interest recently for photonic applications because they possess various unique optical properties and offer great potential for miniaturizing devices. However, with these nanowires, it is not easy to realize tight light confinement or efficient coupling with photonic circuits. Here we show that a high $Q$ nanocavity can be created by placing a single III/V semiconductor nanowire with the diameter less than 100 nm in a grooved waveguide in a Si photonic crystal, and employing nanoprobe manipulation. We have observed very fast spontaneous emission (91 ps) from nanowires accelerated by the strong Purcell enhancement in nanocavities, which proves that unprecedented strong light confinement can be achieved in nanowires. Furthermore, this unique system enables us to move the nanocavity anywhere along the waveguide. This configuration provides us tremendous flexibility in integrated photonics because we can add and displace various functionalities of III/V nanocavity devices in Si photonic circuits.
\end{abstract}

\maketitle

It has been well established that various semiconductor nanowires (NWs) with diameters smaller than 100 nm can be grown by epitaxial methods, such as the vapor-liquid-solid (VLS) method \cite{Wagner1964}. A variety of structures, including core-shell \cite{LauhonLieber2002}, multi-layer heterostructures \cite{GudiksenLieber2002,Tateno2012,Heiss2012}, and p-i-n junctions \cite{DuanLieber2001,WangLieber2001}, have been implemented specifically for III/V semiconductor NWs by appropriately arranging the growth sequence. Several interesting NW photonic devices have been reported, such as lasers \cite{HuangYang2001,DuanLieber2003,Oulton2009Nature,LuYujung2012}, light-emitters \cite{DuanLieber2001}, single photon source \cite{Claudon2010}, photo-detectors \cite{WangLieber2001}, and wavelength converters \cite{NakayamaYang2007}. The dimensions of the NWs naturally suggest that NW-based photonic devices are promising for ultrasmall and power-efficient device operation \cite{YangReview2009}. However, their performance has been limited because no effective light confinement has been achieved for NWs when the NW dimensions are in the sub-wavelength regime. {NWs coupled with plasmonic waveguide modes have been reported \cite{Oulton2007NaturePhoton,Oulton2009Nature}, but the quality factor ($Q$) is rather low ($<$10) and the mode size is smaller than (thus, not compatible with) the NW dimension. Besides the plasmonics, some studies have reported NWs coupled with relatively-small resonators or photonic crystals \cite{Barrelet2006}}, but the light confinement volume is much larger than the wavelength of light and the NWs are too small to achieve a sufficient overlap with the confined mode. This means that we cannot enjoy the various enhancements which would arise from strong light confinement in a tiny active material.

In this report, we propose and demonstrate the design shown in Fig. \ref{Fig1}a in which a sub-wavelength III/V NW is placed in a grooved waveguide in a Si photonic crystal. It has been previously shown that a small local modulation of a line defect waveguide produces high-$Q$ nanocavity modes resulting from the spatial modulation of the mode gap frequency in the line defect. This is a modulated mode-gap cavity, which exhibits a $Q$ of millions and an effective mode volume ($V$) of $\sim(\lambda/n)^{3}$ for well-optimized structures, which $\lambda$ and $n$ are the cavity wavelength and the medium index, respectively \cite{Notomi2010}. It is noteworthy that a surprisingly small modulation can lead to ultra-strong light confinement in this type of structure \cite{Notomi2008,Yokoo2011}. In Fig. \ref{Fig1}a, the sub-wavelength NW produces a certain refractive-index modulation in the line defect, and we confirmed numerically that this modulation is sufficient to form a high-$Q$ nanocavity mode exactly at the NW position \cite{Birowosuto2012b}. As we show later, the cavity mode is mostly localized in the NW. Thus, it can solve the aforementioned problem for NW-based photonic devices. But this is not the only merit of this design. It also enables us to implement various III/V-semiconductor-based nanocavity devices in a Si waveguide platform. Si photonics is now being extensively studied to provide a future photonic platform for integrated circuits, but it suffers severely from the poor optical functionality of Si \cite{Silicon4,Silicon5}. With our proposed design, we can add various III/V functional nanocavity devices at arbitrary positions in Si waveguides. From another viewpoint, NW photonic devices generally suffer from poor input/output coupling \cite{DuanLieber2001,WangLieber2001,HuangYang2001,DuanLieber2003,NakayamaYang2007,Oulton2009Nature,LuYujung2012}, but with our present configuration the NW cavity mode can be easily coupled to Si photonics waveguides as already demonstrated for a simple waveguide \cite{Park2008}.

Moreover, this NW cavity is movable, which is exceptional as regards high-$Q$ nanocavities. Since strong light confinement with a small loss is only achieved by using a special arrangement of photonic crystal cavities \cite{Painter1999,Noda2000} or strongly disordered media \cite{Topolancik2007a,Sapienza2010}, wavelength-sized high-$Q$ nanocavities are prefixed to the surrounding arrangement and are immovable. In our present design, the cavity is created by the perturbation from the NW and is not fixed to the surrounding photonic crystal lattice. {Although reconfigurable cavities have been proposed and reported in various forms in photonic crystals \cite{Yokoo2011,Gardin2008,Smith2008,Seo2009}, but none of them demonstrated the position manipulation of the cavity as a result of moving the single nano-object scatterer (NW). Also with our method, an active material can be embedded in the scatterer so that a nanocavity with active function becomes movable.} This feature adds more flexibility and tunability to the application for Si photonics, and more importantly it may open up an interesting application when combined with a microfluidic circuit \cite{Psaltis2006}. If we fill the groove with a certain liquid, a NW can flow along the groove. Consequently, we may be able to deliver an active NW nanocavity to another location in a microfluidic Si photonic circuit.

In this study, we demonstrate the formation and movability of a cavity by manipulating InAsP/InP NWs on Si photonic crystals with a scanning probe, in other words, by atomic force microscopy (AFM) manipulation. Here we also report the clear observation of a large Purcell enhancement of spontaneous emission from NWs.

First of all, we describe the procedure for making the proposed structure (shown in Fig. \ref{Fig1}a) by AFM manipulation \cite{Junno1995,Benson2010}. After we measured the emission intensities of a number of NWs, we chose some of the NWs for our experiment. Then, we transferred these NWs from the outside of the photonic crystal using an AFM tip with a scanning velocity of 50 nm/s, shown as process (1) in Fig. \ref{Fig1}a. The AFM manipulation technique has been used to slightly adjust the position of nanoparticles on the photonic crystal cavity \cite{Benson2011}, but now we use AFM for manipulating much larger semiconductor NWs over a much larger distance. In fact, the longer dimension of NW is advantageous for long-distance transfer because the NWs will not fall down into photonic crystal holes. By manipulating the AFM tip, we put the NW into the square-grooved waveguide of the photonic crystal, shown as process (2) in Fig. \ref{Fig1}a. Process (3) corresponds to the movement of the NW along the groove that we describe later. Note that the existence of the groove is a great help as regards the controllability of AFM manipulation, and also enlarges the light confinement as shown below.

The electric field intensity profile for the resonant mode calculated by three-dimensional finite difference time domain (3D FDTD) is shown in Fig. \ref{Fig1}a, where we assumed a photonic crystal with a lattice constant $a$ = 352.5 nm and a square groove along the waveguide with a depth $d_{groove}$ of 75 nm and a width $w_{groove}$ of 150 nm. The radius of the hole $r$, the slab thickness $h$, and the width of the waveguide $W$ are 100 nm, 200 nm, and 0.98$\sqrt{3}a$, respectively. We modelled a square-cross-section NW with a length $L_{NW}$ = 2 $\mu$m and a side length $d_{NW}$ = 90 nm (see also Supplementary Fig. S1). These parameters are close to those used in our experiment. Fig. \ref{Fig1}b shows simulated results for the NW on a normal waveguide without a groove. It is apparent that tight light confinement is achieved around the NW after the NW placement. {When we place the NW in the groove, the NW has better overlap with the confined field and the maximum intensity point is moved to the inside of the NW (see Supplementary Fig. S2). Using the fraction of the area of the confined field inside the NW in comparison with the total field, $\sim$ 20 - 30 $\%$ of photons are coupled to the NW for the NW inside the groove while $\lesssim$1$\%$ of photons for the NW placed on the waveguide.} The calculated $Q$ and $V$ values (normalized by $(\lambda/n_{NW})^3$ where $n_{NW}$ is the NW index) for the cavity formed by the NW inside the groove are 33,200 and 0.95, respectively (see Methods). When we compare NWs of the same size on the top and at the center of the photonic crystal waveguide, the $Q$ and $V$ of the former are three-fold larger and four-fold smaller, respectively (see Supplementary Fig. S1). {Our previous simulation \cite{Birowosuto2012b} showed that $d_{NW}$, $w_{groove}$, and $d_{groove}$ should be optimized for increasing $Q$. In addition, $d_{NW}$ should be approximately equal to $d_{groove}$.

Generally speaking, this type of cavities (modulated modegap cavities) are not so sensitive to the detailed shape of the modulation. The overall effective index modulation determines the cavity confinement \cite{Notomi2008}. In fact, it was shown that random disorder in mode-gap waveguides can also create an ultrahigh-$Q$ localized modes \cite{Topolancik2007a,Sapienza2010}, which indicates that an irregular shape of NW does not necessarily lead to low $Q$. From our simulation, the NW diameter influences $Q$ greatly, especially when it becomes closer to the groove width (see Supplementary Fig. S3 and Ref.  \cite{Birowosuto2012b}). We regard that this is more critical issue for creating higher $Q$.}

For our experiments, we fabricated a series of grooved line defect waveguides in Si photonic crystals. We used InAsP/InP heterostructure NWs with ten InAsP quantum disk (QD) layers that worked as emitters \cite{Tateno2012} as illustrated in Fig. \ref{Fig1}c. A scanning electron micrograph of these NWs in Fig. \ref{Fig1}d shows that $L_{NW}$ and $d_{NW}$ are homogeneous and $L_{NW}$ for this sample is about 2 $\mu$m. The InAsP QDs inside a single InP NW without a cap are shown in the transmission electron micrograph in Fig. \ref{Fig1}e. The details of the fabrication and structure are described in Methods and Supplementary Information.

{ In this manuscript, we use sample A (sample B) to refer to the NW with $L_{NW}$ = 1,760 nm (940 nm) and $d_{NW}$ = 83 nm (95 nm) in a photonic crystal with $a$ = 352.5 nm (350 nm).} Fig. \ref{Fig2}a shows an AFM image of the NW after we manipulated the NW inside the groove for sample A (see Supplementary Fig. S4 for the microscope images). This result shows that the NW was successfully placed in the middle of the grooved waveguide. The top of the NW is slightly above the top surface of the photonic crystal and slightly inclined (see Supplementary Fig. S5).

We can locate the NW position directly from the bright photoluminescence (PL) image of InAsP QDs inside the NW at room temperature (RT) using a highly sensitive infrared InGaAs camera (see Fig. \ref{Fig2}b). To confirm the NW cavity formation, we investigated the PL spectra of the same NW on sample A before and after we placed the NW inside the groove (Fig. \ref{Fig2}c). The PL spectrum (black lines) for a single NW of sample A outside the photonic crystal on a bare Si on insulator (SOI) exhibits peaks between 1,200-1,500 nm.

In the PL spectrum (red lines) of the NW inside the groove in Fig. \ref{Fig2}c, we observed a distinctive peak at 1,286 nm. Fig. \ref{Fig2}d shows the magnified PL spectrum recorded with higher resolution, which exhibits a strong-intensity peak at 1,286 and another weak-intensity peak at 1,281 nm. From a comparison with the FDTD calculation, we confirm that the strong-intensity peak with $Q_{exp}^{f}$ = 7,100 is related to the fundamental mode of the cavity shown in Fig. 1a while the latter with $Q_{exp}^{h}$ = 7,500 is related to the second-order cavity mode.  Next, using the emission intensity filtered at 1286 nm (at the resonance peak), we performed a spatially resolved PL scan over the entire area of photonic crystals (as shown in Fig. \ref{Fig2}e) (see also Methods). This result confirmed that the intense emission at 1,286 nm is highly localized at exactly the position of the manipulated NW (see Supplementary Fig. S4 for the microscope image as a position comparison). All of these results proved that the NW placement created a confined resonance mode at the NW position.

Other information regarding the origin of the peak emission can be obtained with a polarization measurement. In Fig. \ref{Fig2}f, we analyze the polarization properties of the sample A emission. When the NW is on a bare SOI, the polarization is parallel to the NW axis ($\theta = 0$). This polarization is expected for a zincblende InAsP QD inside a single InP NW \cite{SasakuraZwiller2012} but this is also related with the different dielectric, the thickness, and the quantum effect of the QD \cite{Tateno2012} (see Supplementary Information). However, the polarization of the NW inside the grooved waveguide is almost perpendicular ($\theta = 70.3 \pm 8.9^{\circ}$) to the NW. Its degree of polarization $\rho$ is 66 $\%$ and is defined as $\rho = (I_{max}-I_{min})/(I_{max}+I_{min})$ , where $I_{max}$ and $I_{min}$ are the intensities at the maximum and minimum of the polar plot. The observed polarization matches the expected polarization for the cavity mode calculated in Fig. \ref{Fig1}a, which strongly supports the previous assignment of the observed resonant mode. {Note that the observed polarization degree is not perfect as simply expected from the cavity mode. We regard that this imperfect polarization may be due to photons directly radiated into the free space and polarization variation of intrinsic NWs. In fact, the polarization degree slightly varies for different samples as shown in Fig. S6.}

Here, we demonstrate the movability of our NW-induced cavity, which is one of the most important features. To demonstrate the movable cavity, we move the NW inside the groove along the waveguide direction and observe the narrow characteristic peak of the cavity at the new position. In this experiment, we used the AFM manipulation technique in sample B three times: first to locate the NW in the groove, and then to displace it twice. After each displacement, we recorded the AFM images, PL images, PL spectra, and spatially-resolved PL scan images as shown in Fig. \ref{Fig3}. We overlay three AFM images that we obtained after the first, second, and third manipulations in Fig. \ref{Fig3}a. The displacement by the second and third manipulations was 3.0, and 6.0 $\mu$m, respectively. We confirmed that the positions of the NW in the AFM images are the same as the bright spots of the NW emission in the PL images of Fig. \ref{Fig3}b. In all three cases, we successfully observed a narrow cavity peak in the PL spectra at a wavelength of $\sim$1,277 nm with slightly different $Q_{exp}$ (5,200 - 2,900 - 3,200) (see blue curves in Fig. \ref{Fig3}c and Supplementary Fig. S3 for the FDTD simulation). At each step, we measured the PL spectrum at a reference NW outside the groove, which showed no difference (green curves in Fig. \ref{Fig3}c). Next, we performed a spatially-resolved PL scan with a filter centered (2-nm window) at the peak wavelength in Fig. \ref{Fig3}c resulting in the PL maps shown in Fig. \ref{Fig3}d. In each image, a single bright peak (in a blue circle) is seen exactly at the NW location shown in Fig. \ref{Fig3}a. The weak spot in a green circle corresponds to the unmanipulated NW we used for the reference. These results clearly show that the sharp resonance moved with the NW displacement. This constitutes the first observation of a movable sub-wavelength-sized high $Q$ cavity.

Our target is to create a nanocavity mode concentrated in NW. To demonstrate this feature, we next investigated the emission lifetime since the strong light confinement in a single NW should lead to a large Purcell enhancement. Here we measured the emission decay curves at 4 K using an 800-nm pulse excitation laser and a bandpass filter (BPF) at the cavity wavelength (see Methods). Since at 4 K the nonradiative recombination rates are much smaller than those at RT \cite{Shields2007}, we should be able to see the Purcell effect more clearly. For our sample, the cavity wavelength and $Q_{exp}$ at 4 K do not differ significantly from the values at RT (see Supplementary Fig. S7). In Fig. \ref{Fig4}a, the emission decay curve of the NW on a bare SOI is shown by black dots. The data can be well fitted with a single exponential curve resulting in an emission lifetime ($1/\Gamma_{NW}$) of 770 ps. Other NWs measured on the bare SOI also show similar values around 800 ps (see Supplementary Fig. S7). In contrast, the NW inside the groove for sample A (red dots in Fig. \ref{Fig4}a) shows distinctive shortening of the PL decay yielding a lifetime ($1/\Gamma_{cav}$) of 187 ps. From a comparison with the emission lifetime of a single NW on a bare SOI, we obtain a 4-fold reduction in the lifetime.

The above result was obtained with a PL decay measurement for different NWs, but we can directly investigate the Purcell effect by measuring the same NW while varying the cavity $Q_{exp}$ by AFM manipulation. In fact, the result in Fig. \ref{Fig3} shows that the cavity $Q_{exp}$ changes at each displacement step (5,200 - 2,900 - 3,200), although the resonant wavelength is mostly the same. We consider that this was because the vertical position of the NW in the groove was changed a little, in other words, part of the NW was lifted up slightly. We confirmed this speculation by pushing the NW down into the groove with the probe after the third displacement, and observed that $Q_{exp}$ recovered from 3,200 to 4,200 (see Supplementary Fig. S8). Our calculation shows that such misplacement leads to a slight reduction in $Q$, but has negligible influence on the cavity volume (see Supplementary Fig. S9). Thus, this manipulation method is ideally suited for Purcell effect characterization. We performed time-resolved emission measurements for sample B at each step of the manipulation shown in Fig. \ref{Fig3}. For the NW on the bare SOI of sample B, the lifetime ($1/\Gamma_{NW}$) was 730 ps (black dots in Fig. \ref{Fig4}b). The PL decay curves of the NW in the grooved waveguide at each step are shown in Fig. \ref{Fig4}b with the corresponding $Q_{exp}$ values. The graph clearly shows that the emission becomes faster as the $Q_{exp}$ increases. At the highest $Q_{exp}$ (5,200), the lifetime was as short as 91 ps. This lifetime is the shortest lifetime reported for the III-V NW. Since this measurement was performed for the same NW in the cavity with the same volume, it represents a clean demonstration of Purcell enhancement, which was realized by the unique feature of our NW-induced nanocavity. For a better comparison, we collect $\Gamma_{cav}/\Gamma_{NW}$ in Fig. \ref{Fig4}c from samples A and B for different $Q_{exp}/V$ values. Note that the $V$ values for samples A and B are 0.95 and 0.54, respectively. {We also added other data points from different NWs of C and D (see Supplementary Figure S10). Sample C (three samples of D) refers to the NW with $L_{NW}$ = 2,620 nm (1,700, 1,800, and 1,900 nm) and $d_{NW}$ = 85 nm (102, 129, and 90 nm) in a photonic crystal with $a$ = 416 nm (382.5, 360, and 360 nm). NW in sample C has a similar structure as A and B while NWs in sample D are InAsP NWs capped with InP layer.} The result clearly reveals a linear relationship between $Q_{exp}/V$ and $\Gamma_{cav}/\Gamma_{NW}$, which is an evidence of the Purcell enhancement.

{Although the inhibited spontaneous emission in NWs has been reported by many groups\cite{Bleuse2011,Bulgarini2012}, there has been very little work on Purcell enhancement observation for NWs. Recently, Oulton et al. observed a six-fold shortening of the PL lifetime in a single CdS NW coupled to plasmonic waveguides (no cavity effect)\cite{Oulton2009Nature}. Since these plasmonic structures realize an extremely small mode volume, it can lead to large Purcell enhancement even with very low $Q$ resonance. Our present result gives the clear Purcell enhancement of a single NW for the first time in a high-$Q$ dielectric cavity. The achieved enhancement is slightly larger than the result in the NW-plasmonic system. Although there are pros and cons for both methods, we regard that our NW system is advantageous for integrated photonics applications because of its low loss and ease of integration with the waveguide. Although the confinement factor (15 - 20 $\%$) is comparable to that for our cavities ($\sim$ 20 - 30 $\%$), the plasmonic mode is much smaller than the NW dimension, which generally degrades the energy efficiency for device operations. Furthermore, extreme plasmonic modes may suffer from nonradiative energy transfer to metals\cite{FordWeber1984}, which makes the observation of large Purcell enhancement difficult. In contrast, our system may still have a room for achieving larger enhancement if we boost up the cavity $Q$ as we discuss below.} 

The theoretical Purcell factor $F_{P}$ for our sample with the fastest emission rate is about 732. As is well known, the emission rate will be reduced for several reasons, such as a spectral, spatial, or polarization mismatch \cite{Ryu2003,Englund2005}. Note that in our experiments, the polarization mismatch is large (as seen in Fig. \ref{Fig2}f). A quantitative investigation of these reduction factors is beyond the scope of this study. Although we could not determine the exact position of the emitter that we measured in this experiment, our AFM manipulation technique is able to precisely align the emitter position at the field maximum of the cavity if we know the precise position of the emitter in the NW. This would constitute an important advantage of our system over other solid-state cavity QED systems \cite{Badolato2005}.

In conclusion, we demonstrated that a high-$Q$ cavity can be created simply by putting a single NW into a groove in a photonic crystal waveguide. We used the spontaneous emission of the QDs inside the single NW to probe the cavity formation by comparing the PL spectra and the emission polarization of the NW on a bare SOI and inside the groove. We moved the cavity spatially while manipulating the position of the NW. Furthermore, we also observed emission rate enhancements for the NW inside the groove showing unprecedented strong light confinement in a single NW with a record of eight-fold emission rate enhancement. We believe that our observation will stimulate the discovery of new phenomena in fundamental research on cavities in photonic crystals \cite{Painter1999,Noda2000} and also the localized modes in disordered photonic systems \cite{Sapienza2010}. As regards the realization of novel nanophotonic devices, our new approach for creating cavities is useful for movable devices such as nanolasers \cite{Matsuo2010} or all-optical memories \cite{Nozaki2012} with the capability of integrating them on a Si platform. { Especially, a laser is one of the most interesting applications of our system. To achieve lasing,  we may need a larger gain volume and higher $Q$, which remains for a future study. As for the improvement of $Q$, smaller-diameter NWs in narrower grooves are possible options.} Finally, the integration technique of NW is not limited to III-V NWs on Si photonic crystals, e.g. diamond nanowires \cite{Babinec2010} with suitable refractive index photonic crystals. Another potential method for moving cavities, namely nanofluidic technology \cite{Psaltis2006} remains tantalizing.

\section*{Methods}
{\footnotesize \subsection*{Finite difference time domain and Purcell factor}
We performed a 3D FDTD calculation to investigate the field distribution. Some parameters, such as the calculation area, the perfectly matched layer boundaries, and the grid, have already been described elsewhere \cite{Birowosuto2012b}. The strength of the light confinement or $Q$ was calculated from a single-exponential fitting of the energy decay curve tail with the narrow band excitation of the cavity. From the fitting, we determined the cavity photon lifetime $\tau_{ph}$ and we estimated $Q$ in this cavity as $Q = \tau_{ph}\cdot\omega_{c}$.

The common figure of merit of the spontaneous emission rate enhancement in the cavity compared with that in the bulk at resonance is the Purcell factor $F_{p}$, which can be written as follows:
\begin{equation*}
F_{p} = \frac{3Q}{4\pi{}^{2}V}
\end{equation*}
where mode volume $V$ denotes the dimensionless mode volume with a cubic wavelength $\lambda$ of the emission inside a single NW with a refractive index $n_{NW}$. Note that $E(\textbf{r}_{max})$ is located inside the NW.
\begin{equation*}
V = \frac{\int{\epsilon(\textbf{r})|E(\textbf{r})|^{2}d^{3}r}}{\epsilon(\textbf{r}_{max})|E(\textbf{r}_{max})|^{2}}\left(\frac{n_{NW}}{\lambda}\right)^{3}
\end{equation*}

\subsection*{Photonic crystal fabrication}
We fabricated a series of Si photonic crystals with a groove in the middle of the waveguide by electron-beam lithography and inductively coupled plasma etching. Holes with a diameter of 95 $\pm$ 5 nm were fabricated. The underlying SiO$_{2}$ layer were then removed by selective wet etching using HF solution. This technology guarantees a resolution of $ < 5 \%$ in diameter and $ < 1 \%$ in distance between the holes and all these processes resulted a photonic crystal slab with a thickness of 200 nm. Finally, the groove was fabricated after a second-mask process. We obtained this desired groove depth after determining a suitable etching time through the linear dependence of the groove depth on variations in the etching time.

\subsection*{NW growth and transfer}
The NW growth was carried out in a low-pressure horizontal metalorganic vapor phase epitaxy (MOVPE) reactor. Trimethylindium (TMIn) was the group III sources while phosphine (PH$_{3}$), tertiary-butyl phosphine (TBP), arsine (AsH$_{3}$), and tertiary-butyl arsine (TBAs) were the group V source. The substrate was InP(111)B and the catalysts were Au particles (20 nm in diameter) obtained from Au colloids. Ten InAsP layers were grown alternately with InP layers in a single NW capped with a 30-nm diameter-thick InP layer. The InAsP layer thickness was 10 nm and the growth time for each InAsP and InP layer was 10 s. The capping growth process was performed at 470$^{\circ}$C for 35 minutes. After the growth process, we obtained a vertically-grown NW with alternating InAsP/InP layers located in the middle and in the core-shell region. { The NWs were dispersed from the grown substrate to the photonic crystal substrates by pushing gently both substrates to each other resulting a minimum number of broken NWs (mechanical dispersion technique)\cite{Tateno2012}.}

\subsection*{Photoluminescence measurements}
We used a standard $\mu$PL setup. The sample was optically excited either using the CW diode laser at 640 nm or picosecond pulses from a 80 MHz mode-locked titanium sapphire laser at 800 nm with an average excitation power of 100 $\mu$W. A free-space excitation technique was applied through a 0.42- and 0.50-numerical-aperture near-infrared microscope objective, by which the beam diameter was estimated to be $\sim$2 $\mu$m. The fraction of the spontaneously emitted photons that coupled to the radiation modes was collected through the same microscope objective and filtered with a dichroic beam splitter and a long-wave-pass filter (LWPF). For direct observation of the emission from a single NW, we split the emission from the NW and the visible illumination light using a dichroic mirror (cut off wavelength 1,064 nm). The detection parts contained a visible-near infrared CCD camera and a highly sensitive short wave infrared InGaAs camera with an LWPF (cut off wavelength 1,200 nm) for the microscope image and the PL image, respectively. To measure the PL spectrum, we coupled the emission into the multimode fiber and directed it to a grating spectrometer with a cooled InGaAs array. It had resolution limits of 0.24 and 0.05 nm at 1,550 nm for 300 grooves/mm and 1,000 grooves/mm, respectively. To obtain the spatial origin of the emission, we also performed a spatially resolved PL scan. We selected the target cavity wavelength and chose the best resolution limit. Finally, we collected PL spectra from a 7.5 $\mu$m x 18.5 $\mu${m} area of photonic crystals with a step of 500 nm and the initial point of the photonic crystal edge (see Supplementary Fig. S2). For the time-resolved PL measurements, we coupled the emission filtered by a band-pass filter of 1,300 $\pm$ 30 nm to a single mode fiber and directed it to one-channel NbN superconducting single photon detector (SSPD). Finally, we connected the SSPD and the excitation pulse signals to one input and a synchronization input channel of a time-correlated single photon counting (TCSPC) board, respectively.}

\section*{Acknowledgments}
We acknowledge S. Fujiura and M. Ono 
for assistance with the NW manipulation.

\section*{Author contributions}
M. D. B., A. Y., G. Z., and M. N. conceived the idea and designed the experiments. M. D. B. performed the simulation, conducted the experiments and analysed the data. M. D. B. and M. N. wrote the manuscript. A. Y. manipulated the nanowires. K. T. conducted the NW growth. G. Z. and E. K. involved in the fabrications. M. T. supported the experiments. H. T. supported the simulation. M. N. guided the project.

\section*{Additional information}
The authors declare that they have no
competing financial interests. Reprints and permission information
is available online at http://npg.nature.com/reprintsandpermissions/.
Correspondence and requests for materials should be addressed to\\
M. D. B. (m.d.birowosuto@gmail.com) and M. N. (notomi.masaya@lab.ntt.co.jp)\\

\newpage

\begin{figure}[htbp]
\caption{{\small \textbf{A movable hybrid cavity and emitter in a photonic crystal and nanowires.} \textbf{a}, Schematic of a movable cavity using a single semiconductor nanowire (NW) inside a square-grooved waveguide in a 2D Si photonic crystal while an atomic force microscope (AFM) tip is manipulating a single NW. Calculated intensity distribution ($\left|\emph{E}\right|^{2}$) is shown as a surface map of the photonic crystal. \textbf{b}, $\left|\emph{E}\right|^{2}$ distributions for a single NW on a simple waveguide. \textbf{c}, Structure of a single InP-capped heterostructure NW with InAsP quantum-disk (QD) emitters. \textbf{d,e}, Scanning electron (d) and transmission electron (e) micrographs of a single NW with a length of $\sim$2 $\mu$m and 10 InAsP QDs inside each NW. The scale bars in (d) and (e) indicate 2 $\mu$m and 50 nm, respectively.}}
\label{Fig1}
\end{figure}

\begin{figure}[htbp]
\caption{{\small \textbf{Manipulation of single NW and photoluminescence (PL) characterizations.} \textbf{a,b}, False-color AFM (a) and PL images (b) of the single NW inside the square-grooved waveguide of the photonic crystal. \textbf{c}, PL spectra of a single NW on a bare Si on insulator (SOI) (black lines) and inside the square-grooved waveguide of the photonic crystal (red lines). \textbf{d}, Details of the cavity spectrum and the green lines show the Lorentzian fits. \textbf{e}, Spatially-resolved PL scan of the cavity at 1286 nm wavelength. The orange and purple lines show the photonic crystal structures, and the AFM image area, respectively. The scale bars in all images indicate 2 $\mu$m. \textbf{f}, Integrated intensity emission of the NW inside the square-grooved waveguide of the photonic crystal (red spheres) and the NW on a bare SOI (blue spheres) as a function of the linear polarization angle at RT and 1286 nm. The green arrow and the solid lines represent the NW orientation and the sinusoidal fits, respectively. All measurements were performed for sample A using continuous-wave (CW) laser excitation at 640 nm and at room temperature (RT).}}
\label{Fig2}
\end{figure}

\begin{figure}[htbp]
\caption{{\small \textbf{Observation of a movable cavity at three different positions.} \textbf{a,b}, False-color image of an overlay from three AFM images (a) and PL images (b) of a single NW at three different positions inside a groove in a photonic crystal. The black rectangles represent the cropped area of the AFM images. \textbf{c}, PL spectra of the cavity and the NW background emission shown by blue and green lines, respectively. \textbf{d}, Spatially-resolved PL scan of the cavity intensity integrated over the 2-nm cavity peak. The blue and green circles correspond to the PL spectra in (c). The orange and purple lines show the photonic crystal structures and the AFM image size, respectively. All measurements were performed for sample B using CW laser excitation at 640 nm and at RT. The scale bar represents 2 $\mu$m.}}
\label{Fig3}
\end{figure}

\begin{figure}[htbp]
\caption{{\small \textbf{Emission rate enhancement for different $Q_{exp}/V$ values.} \textbf{a}, Time-resolved emission of a QD in sample A. \textbf{b}, Time-resolved emission of a QD in sample B with a different position inside the groove in the waveguides. Decay curves for a single NW on a bare SOI outside the photonic crystal are shown as black spheres and those for an NW inside the grooves of samples A and B are shown as red and purple spheres, respectively. The intensity at zero delay of each curve in (a) and (b) is shifted for clarity. The white lines and the green spheres represent a single exponential fit and an instrumental response function of 45 ps. \textbf{c}, Summary of spontaneous emission rate enhancement for different $Q_{exp}/V$ values. {Four samples with different NWs, C and D, were added (see also Supplementary Fig. S10).} The blue dashed line is a linear fit. All measurements were performed using a picosecond-pulse laser emitting at 800 nm and at 4 K.}}
\label{Fig4}
\end{figure}

\clearpage

\centering\includegraphics[width=15cm]{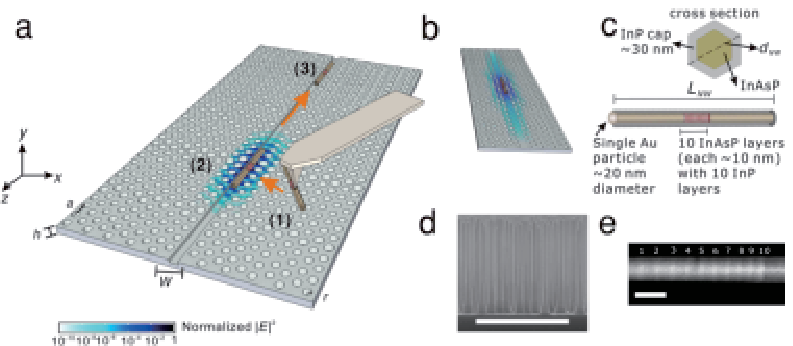}
Fig. 1
\newpage

\centering\includegraphics[width=15cm]{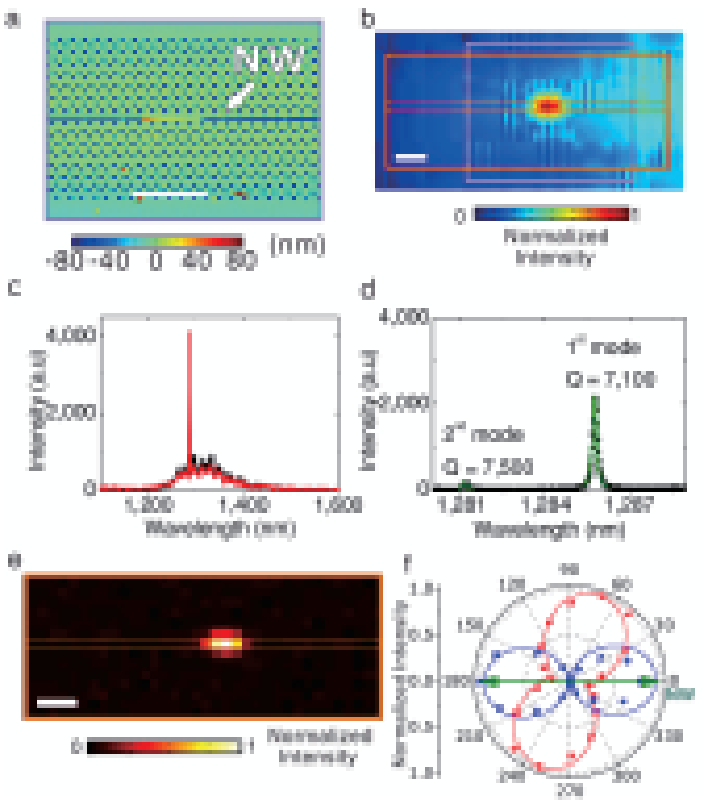}
Fig. 2
\newpage

\centering\includegraphics[width=15cm]{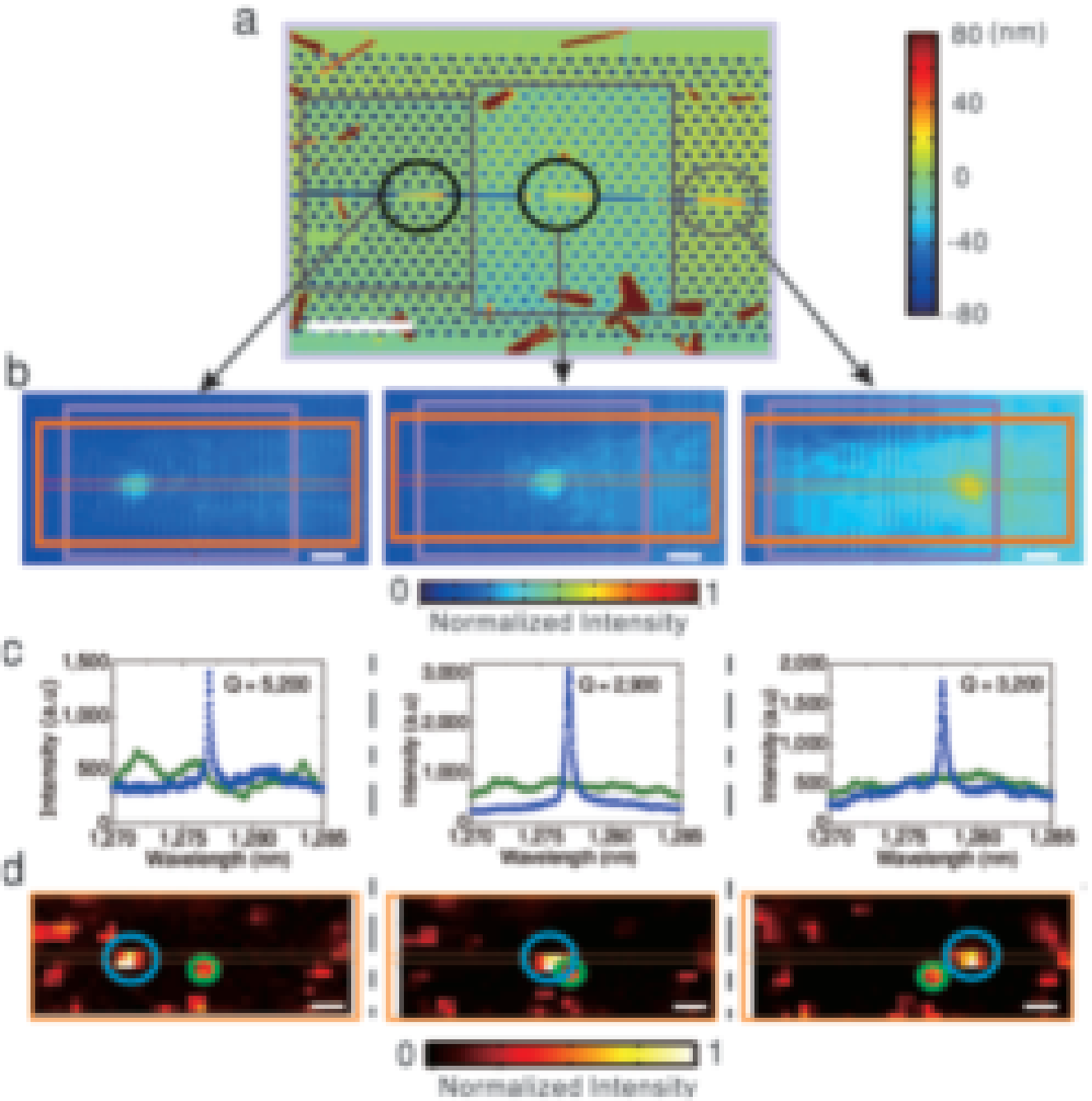}
Fig. 3
\newpage

\centering\includegraphics[width=11cm]{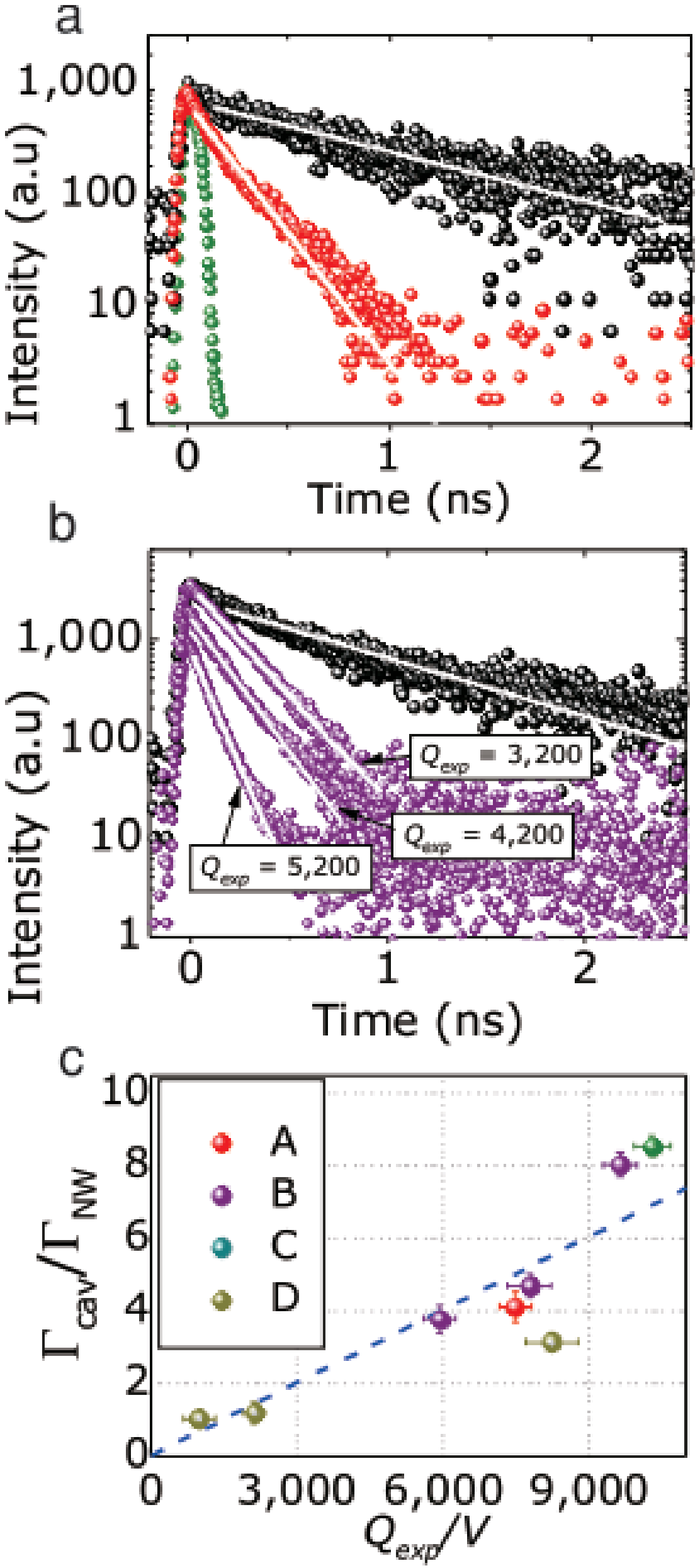}
Fig. 4

\end{document}